# Dark Matter in Zwicky's Cosmology: Towards an Epistemological Reconstruction

Simon Beyne[1 & 2] & Christian Marinoni[2]

## Abstract

A new contextualised reading of Fritz Zwicky's 1933 article "The redshift of extragalactic nebulae" about the virial analysis of the velocity dispersion of galaxies in the Coma cluster leads to a reconsideration of the traditional discourse on the introduction of dark matter. We argue that this component of matter was not only already on the stage of the scientific debates of the time, but also, in a more concealed form, played a central role in Zwicky's epistemic context. We thus reject the narration that dark matter is the result of a ``naïve'' *astrophysical* observation and emphasise the *cosmological* motivations that prompted Zwicky to presciently search for it. Moreover, with regard to its abundance, we argue that the discrepancy between the observed amount of luminous matter in the Coma Cluster and Zwicky's higher mass estimate derived from virial analysis was not, in fact, astonishing. What Zwicky described as a surprising excess of dark matter was of precisely the order of magnitude he had set out to identify. Consequently, we challenge the widespread view that dark matter was merely an *ad hoc* hypothesis introduced to rescue Newtonian theory. Instead, we suggest it may represent one of the earliest cosmological indications supporting a new emerging theory of gravitation: General Relativity. This reinterpretation contributes to ongoing debates in the philosophy of science concerning the epistemic status of *ad hoc* hypotheses.

---

[1] Centre Gilles Gaston Granger - Aix Marseille Univ, CNRS, CGGG, Marseille, France
[2] Centre de Physique Théorique - Aix Marseille Univ, Université de Toulon, CNRS, CPT, Marseille, France

## Introduction

In most literature, both specialized and popular, dark matter is presented as an ad hoc hypothesis introduced, by Fritz Zwicky in 1933 to explain the serendipitous discovery of an unexpectedly large, and therefore anomalous, gravitational field within the Coma cluster of galaxies (Bonnet-Bidaud & Lepeltier, 2021, p.221; Biermann; 2022, p.146; Freeman, 2013, p.63; Horvath, 2021, p.328 ; Hossenfelder & McGaugh, 2018, p.36; Kosso, 2013, p.143 Luminet, 2021, p.1 ; Sadoulet, 1996, p.877 ; Turner, 2000, p.624 ; Van den Bergh, 1999, p.657).

In this paper we deconstruct such a traditional narrative by proposing a new reading of this crucial episode in the history of science which centers around the following theses

1) Zwicky's discovery was not 'naïve' but programmatically prepared, and the scientific motivation behind analyzing the velocity dispersion of galaxies in the Coma cluster was intrinsically cosmological, rather than purely astrophysical or dynamical.
2) Zwicky was not the first scientist to introduce nor exploit the concept of dark matter in an astronomical context.
3) The large velocity dispersion observed by Zwicky in the Coma cluster was not regarded as an 'anomaly,' at least from his perspective.
4) Zwicky dark matter is not an *ad hoc* hypothesis.

Point 1) and 3) are overlooked in the contemporary literature[3]. Stressing that Zwicky's analysis is part of a cosmological research programme, and not an unexpected research accident, allows us to demystify the narrative that Zwicky introduces the notion of dark matter ``to save'' Newtonian mechanics. Point 2) is well known and has already been resumed for example in a recent article by Gianfranco Bertone and Dan Hooper (2016). With point 4), developed beginning in section 5, we argue that the so-called ad hoc character attributed to the dark matter hypothesis does not reflect Zwicky's intentions but is essentially a posteriori, independently rooted in both philosophical and physics communities, and sustained by the lack of a consensus definition of what 'ad hoc' means.

---

[3] It should be noted that Longair (2004, p.10; 2006, p.116) had hypothesised that the need for cosmological dark matter was emphasised by Einstein and de Sitter for their model in 1932. Nevertheless, we make an original contribution: we highlight Zwicky's intentions and trace the presence of dark matter in de Sitter's work.

## 2 The context

To understand the structure and purposes of Zwicky's 1933 article "*Die Rotverschiebung von extragalaktischen Nebeln*", it is necessary to situate his research interests and activities around that time in their proper context. This will allow us to understand the epistemological terrain in which Zwicky's analysis of the Coma cluster is rooted, and to better appreciate the link between this specific study and Zwicky's broader cosmological discourse.

### 2.1 Redshift : an observable for testing cosmological theories

Research into cosmological questions in the years immediately preceding 1933 was what Merleau-Ponty (1965) described as essentially being carried out in parallel by two communities that ignored each other: theoretical physicists and observational astronomers. Only a few physicists had abilities and skills to work at the interface of these two communities, among them Arthur Eddington, Willem de Sitter, and Fritz Zwicky himself. Although they used different methods of analysis, members of both communities eventually came to the same conclusion, that the universe appears to be evolving dynamically. On the one hand, a shift in the spectra of galaxies (the so called redshift effect) was observed and interpreted as a Doppler consequence of the recession velocity of galaxies. On the other hand, a stretch of the metric scale factor was theorised, also called *cosmological redshift*, and was associated with the dilation of physical space. Understanding the physical meaning of the redshift effect was thus one of those hot research topics at the center of intense research and debate in the years around 1930, and a natural polemicist like Zwicky was one of the contenders in the arena. Indeed, beginning with an article written in 1929 (Zwicky, 1929), and throughout his career, he considered, analyzed, adopted, and rejected several possible interpretations of the observed redshift effect.

Zwicky, who had trained as a physicist before becoming an astronomer, always kept abreast of the work of theorists such as Friedman, Tolman and Lemaître. He was aware also of the works of Eddington and of the discovery that the static universe of Einstein was unstable and therefore prone to evolution, a state that seemed to be in better agreement with Hubble and Humason's observational findings. In light of the interpretive difficulties posed by the results of general relativity, the distinction between *cosmological redshift* (dilatation of space between galaxies) and *doppler velocity* (motion of galaxies relative to space) was not yet clearly appreciated at the time. Consequently, Zwicky, like many astronomers, Hubble himself among them, thought it unlikely that redshift (and particularly its large values measured in galaxies) provided an unequivocal indication that the universe was expanding. He therefore went so far as to doubt the soundness of general relativistic models of the universe and to investigate, as it was mainstream activity at the time, "*whether the redshift is due to recession or other causes*" (Hubble & Tolman, 1935). Indeed, according to his biographer: "*In 1931, Humason came up with a velocity of 19,600 kilometres per second (12,179 miles per second) for a galaxy cluster in the constellation Leo. These enormous speeds convinced Zwicky that the theory was dead wrong. He was hardly alone at the time. James Jeans dismissed the breathtaking speeds being found as impossible. Zwicky immediately set about trying to figure out why the measurements being made by Hubble and Humason didn't mean what they appeared to mean: that the universe was blowing up.*" (Johnson, 2019, p.68). Later on, in 1971, Zwicky also explained in an interview that he could not accept "*all of the blah blahs of the expanding universe*" [4].

This highly critical and skeptical stance toward the relativistic interpretation of redshift should not be taken as evidence that Zwicky was biased against Einstein's general theory of relativity. While, on the one hand, he actively sought to identify potential limitations in its theoretical foundations and empirical predictions, it is important to emphasize that, on the other hand, he was equally committed to exploring the predictive power of the emerging theory and to seeking observational evidence for its cosmological implications. For example, Zwicky was the first to recognize the cosmological significance of gravitational lensing, an astonishing prediction of Einstein's theory. While Einstein's calculations indicated that, within the Galaxy, a star located along the line of sight between a source and an observer would deflect light only minimally—rendering the effect astronomically negligible —Zwicky demonstrated that if both the emitting source and the deflecting lens were entire galaxies in intergalactic space, the deflection could be substantial and potentially observable with telescopes. He further speculated that this phenomenon could, in principle, be used to determine galactic masses, potentially with even greater precision than afforded by the virial theorem. "*The problem of determining nebular masses at present has arrived at a stalemate…… Observations on the deflection of light around nebulae may provide the most direct determination of nebular masses and clear up the above-mentioned discrepancy.*" (Zwicky, 1937, p.290).

---

[4] Interview with R. Cargill Hall, historian, Jet Propulsion Laboratory, 17 May, 1971.

Pursuing an original line of inquiry, Zwicky proposed the *tired light* hypothesis as an alternative explanation for the observed linear relation between cosmological redshift and distance. In this framework, first advanced in 1929, redshift arises from a gradual loss of photon momentum during propagation through intergalactic space, as energy is transferred to matter: "*loses a certain amount of impulsion on the way from P1 to P2 and gives it to matter. The photon becomes redder*" (Zwicky, 1933a, p.123). Zwicky interpreted this mechanism as a gravitational analogue of the electromagnetic Compton effect and presented preliminary estimates suggesting that the resulting redshift would be proportional not only to distance but also to the density of matter in intergalactic space.

With respect to this latter cosmological parameter, he stated: "*For the total space investigated, the possible limits for ρ are, according to E. Hubble, $10^{-26}$ g/cm³ > ρ > $10^{-31}$ g/cm³.*" (1929, p. 778) Notably, only the upper bound of this density interval was consistent with Zwicky's own interpretation of the redshift. The large redshifts measured by Hubble could be reproduced if the density approached this upper limit. In pursuing an alternative interpretation of the cosmological redshift—distinct from and antagonist to that offered by general relativity, namely the stretching of the spatial metric of the universe—Zwicky, in our view, appears to have already formed a strong bias by 1929 regarding which end of the interval for Hubble's density estimates was of greater "cosmological" significance.

We thus emphasize that, as early as 1929, Zwicky recognized the enormous potential of this new observable. He conceived the redshift as a novel epistemic tool for assessing the validity of cosmological theories—an observable with which every model would necessarily have to contend: "*It would be very important to measure the radial velocities of as many globular clusters as possible in order to decide definitively between the different theories. It is particularly desirable to determine the redshift independently of the proper velocities of the observed objects*" (Zwicky 1929, p.778) and to learn more about the physics of the universe : "*It is also pointed out that an extensive investigation of great clusters of nebulae will furnish us with decisive information regarding the question whether physical conditions in the known parts of the universe are merely fluctuating around a stationary state or whether they are continually and systematically changing*" (Zwicky, 1937, p.218). The 1933 article *"The Redshift of Extragalactic Nebulae"*, as its title indicates, represents thus the outcome of a long period of intellectual development and was specifically conceived with two principal aims: the ontological task of clarifying the nature of the redshift, and the heuristic task of exploring its cosmological utility.[5]

## 2.2 – Dark matter: an *inevitable* component of the universe

The idea that many celestial bodies do not appear luminous is not new. It could be traced back to the late 18th century, when John Mitchell and Pierre-Simon de Laplace independently theorised about the existence of ``occluded stars"[6] : "It is therefore possible that the largest luminous bodies in the universe may, by this cause, be invisible." (Laplace, 1798). This is not dark matter, in the sense of matter that does not emit electromagnetic waves, since this hypothesis refers to objects that emit particles of light but have such an intense gravitational field that they can reabsorb it. For the first time, however, the combined application, in an astronomical context, of established physical theories—such as Newtonian dynamics—and speculative hypotheses—to such as the massive-particle nature of light—facilitated the emergence of the notion of invisible matter characterized by strong gravitational fields.

At the turn of the nineteenth century, prior to the advent of General Relativity in astronomy, another intrinsically interdisciplinary analysis was emerging at the intersection of physics and astronomy. This new line of inquiry explored the possibility of applying concepts and mathematical formalisms from the kinetic theory of gases to the study of self-gravitating astronomical systems, such as the Milky Way. These investigations ultimately led to the development of tools—or, more precisely, "ideal balances"—designed to quantify the total mass of the universe, thereby allowing for the inclusion of possible contributions from non-luminous matter.

Lord Kelvin was the first, in 1904, to propose that the virial theorem—originally derived by Clausius in 1870 within the framework of the kinetic theory of gases—could be applied beyond the confines of laboratory physics to astronomical systems on galactic scales. Specifically, Kelvin suggested its use for studying the structure of the Milky Way and, in particular, for estimating the total mass of a gravitationally bound system, rather than only the luminous, and thus directly visible, component. This conceptual shift opened a new astronomical window for the detection of non-luminous matter in the universe, variously described in the early literature as *"dark bodies"* (Kelvin 1904, p. 277), *"matière obscure"* (Poincaré 1906, p. 158), *"dark matter"* (Kapteyn 1922, p. 302), and *"invisible mass"* (Oort 1932), before Zwicky's designation *"dunkle materie"* in 1933. And indeed the existence of a significant fraction of dark matter was also demonstrated (e.g. Oort 1927a) . For an in-depth historical reconstruction of the genesis and evolution of the concept of dark matter up to the pivotal year 1933, see Bertone and Hooper (2016). See also the synthesis of historical references to dark matter between 1844 and 1982 presented in Table 1 of Ben-Amots (2021, p. 17).

---

[5] The *redshift* would make it possible to determine the distances at which cosmic objects are found, to describe their dynamics, to develop cosmological models, to understand the relation between light and matter and to interpret cosmic rays; each of these aspects is the subject of a part or sub-part of the 1933 article.
[6] Stars so massive, they could not let out light. John Mitchell had previously suggested the existence of such bodies, '*dark stars*', and even proposed to detect them by their gravitational effects.

Obviously, in the sense understood by the first astronomers who addressed the problem, the term *dark matter* did not refer to the exotic, non-baryonic component postulated in the current standard cosmological model. Rather, it denoted the discrepancy between the mass theoretically inferred from the dynamics of gravitationally bound systems and the luminous, or "shining," mass detected through telescopic observations. In this historical usage, dark matter referred to material that emitted only faint amounts of radiation—such as dim or obscured objects—or none at all. With the current understanding, this category would include gas clouds, brown dwarfs, white dwarfs, neutron stars, black holes, and similar bodies.

These arguments demonstrate that the expression *dark matter* was not originally introduced by Zwicky as a hypothesis to account for his observations, rather, it already occupied a place within the established astronomical lexicon. This stands in contrast to claims that continue to surface in the literature—not only in non-specialist or popular-science venues[7], but also within academic contexts. Such assertions appear in astronomical research (e.g. Porter, Johnson & Graham, 2011), philosophical discussions (e.g. Allzén, 2021), and even historical treatments in monographs (e.g., Sanders, 2010, p.13). For example, Sanders explicitly remarks: "Unquestionably, Fritz Zwicky was the first to propose this form of dark matter."

When, in 1933, Zwicky extended Kelvin's earlier idea by modelling not the Milky Way and its stellar content, but instead the Coma cluster and its galaxies as a self-gravitating gas, his explicit aim was to determine the fraction of the cluster's total mass that was dark—an estimate which, in his view, could also serve as a reliable proxy for the cosmic abundance of dark matter (as we will see later in the article). More broadly, when examined within its historical context, the assumption of the existence of dark matter appears far less exotic or ad hoc than it is sometimes portrayed. On the contrary, it emerges as an almost inevitable hypothesis—considerably less audacious than the alternative presumption that all cosmic matter is luminous.

---

[7] E.g. https://www.naturalhistorymag.com/features/011330/dark-matter

## 3 Theoretical Necessity of Dark Matter

One significant development, as we discussed, was the establishment of a new observational approach through the application of the virial theorem in astronomy, enabling the estimation of the fraction of matter not accounted for by optical surveys of the universe. Equally important is the examination of whether a dark component was theoretically required in contemporary models—that is, whether, prior to the pivotal year 1933, any cosmological or astrophysical framework necessitated or postulated the existence of substantial amounts of dark matter to remain viable.

### 3.1 The case of cosmic rays

Zwicky's paper, *"How Far Do Cosmic Rays Travel"* (1933b), constitutes an important document in understanding the early role that the notion of dark matter played in his cosmological framework even before he obtained a robust estimate of its abundance through the analysis of the velocity dispersion of galaxies in the Coma cluster. Notably, in this work, published in January 1933 before his paper on Coma, Zwicky advanced a tentative indirect argument for the existence of dark matter.

At the time, cosmic rays were known to be extremely energetic and penetrating, but their origin and propagation range were unclear. The earliest studies of cosmic rays, pioneered by Millikan, showed that they arrive at Earth from all directions – rather than preferentially from the Sun or our Galaxy (Millikan 1926, 1928) – a result that supported the idea of their non-local, possibly cosmological origin.

Zwicky, who had served as Millikan's assistant at Caltech some years earlier, was among the first to recognize that cosmic rays could provide a new window onto the large-scale distribution of matter in the universe. In his 1933(b) paper, written with the explicit aim of assessing the plausibility of Millikan's claims, Zwicky postulated that cosmic rays are produced in amounts proportional to the density of matter in the universe and that, during their propagation to the Earth, they undergo energy losses, scattering, and absorption through interactions with matter. By doing so, he effectively shifted the problem from the properties of cosmic rays themselves to the physical characteristics of intergalactic matter. In detail Zwicky advanced the idea that the very existence of cosmic rays of cosmological origin could provide indirect evidence for a substantial quantity of matter – possibly dark matter (in Zwicky's own terms) – pervading intergalactic space.

Since the ratio of light emitted by the Milky Way to that emitted by the rest of the universe is much greater than unity – whereas the corresponding ratio for cosmic rays is much smaller than unity – Zwicky argued that cosmic rays are unlikely to be produced by luminous matter alone and must instead originate in invisible matter distributed throughout intergalactic space: in his own words "*it is therefore impossible for cosmic rays, if they are photons, to come from luminous matter*" (1933b, p.147). Zwicky then extrapolated the linear redshift–distance relation to arbitrarily large distances and, more critically, assumed that cosmic rays are produced at a rate proportional to the density of matter in intergalactic space. On the basis of empirical estimates – according to which "*the average dark matter density in our galaxy ($\rho_g$) and in the whole rest of the universe ($\rho_u$) are in the ratio $\rho_g/\rho_u > 100,000$*" (p.147) he finally concluded that cosmic rays would have to be generated by matter distributed over a region of the universe whose size greatly exceeds (by two order of magnitude) the maximum distance allowed by the redshift–distance relation once energy losses during particle propagation are taken into account. In other terms, if intergalactic space were too empty, cosmic rays would travel essentially unimpeded over unphysical distances.

One possible way to alleviate this theoretical difficulty and preserve Millikan's proposal was to speculate that the linearity of the redshift–distance relation breaks down at large distances, although Zwicky did not explicitly invoke his 1929 tired-light model in this context (see Section 2.1). More importantly for our purposes, however, he also outlined an alternative resolution: namely, accepting that the ratio $\rho_9/\rho_u$ is much smaller than traditionally assumed. This, in turn, would imply – among other consequences – "*contradicting the so far emptiness of extragalactic space*" (p. 147).

Zwicky neither specified any quantitative limit on the matter density required to 'save the phenomenon' nor accorded any privileged status to this particular solution relative to the other possible resolutions he discussed. For this reason the work cannot straightforwardly be classified as a discovery paper. Nevertheless, it is highly instructive in that it develops an interesting indirect argument for the existence of invisible matter that is entirely independent of dynamical considerations derived from the motions of cosmic objects. Instead, the argument entirely rests on the modeling and interpretation of the propagation and interaction of cosmic particles, such as gamma rays. Most importantly for the present analysis, this investigation of cosmic rays indicates that, even before the publication of his papers on the Coma cluster, Zwicky was already disposed to entertain the possible existence of substantial amounts of unseen matter distributed throughout cosmic space beyond galaxies, and was likely seeking independent confirmation of its presence. Zwicky would change his view of the mechanism that produces cosmic rays only a year later. In a paper written together with Baade in 1934 he realises that, on closer examination, the hypothesis that cosmic rays originate in intergalactic space implies that "*one is forced to assume entirely fantastic processes as regards the mode of creation of the rays*" (Baade & Zwicky, 1934). He therefore put forward an entirely new proposal, with the aim of resolving some of the major difficulties concerning the origin of cosmic rays. Accordingly, the production of cosmic rays is not proportional to the density of intergalactic matter but linked to a sporadic and transient process, such as the flare-up of a supernovae.

*3.2 The case of the Einstein de Sitter cosmological model*

The year before Zwicky's paper on the analysis of the Coma cluster, Einstein and de Sitter (EdS) obtained a non-static cosmological solution of the general relativistic field equations, consistent with the cosmological principle and capable in principle of providing a theoretical framework for interpreting Hubble's 1929 discovery of the linear relation between galactic redshift and distance. This solution was also appealing from a theoretical standpoint, as it constituted the first dynamical model of the universe formulated without the need for either a cosmological constant or spatial curvature.

This model is characterized by zero spatial curvature, corresponding to a universe whose spatial sections are flat, Euclidean three-dimensional hypersurfaces. Since the Einstein field equations establish a link between the geometry of spacetime and its matter/energy content, a measurement of the abundance of cosmic matter (more properly of its spatial density) is enough, in principle, to tell apart relativistic models with different curvatures (positive, negative or null). In 1932, however, there were still no precise cosmological observations that could be instrumental to this model selection purpose. As Einstein and de Sitter wrote in their paper:

> *"There is no direct observational evidence for the curvature, the only directly observed data being the mean density and the expansion, which latter proves that the actual universe corresponds to the non-statical case. It is therefore clear that from the direct data of observation we can derive neither the sign nor the value of the curvature, and the question arises whether it is possible to represent the observed facts without introducing a curvature at all."* (Einstein & de Sitter, 1932, p.213)

In the absence of empirical evidence favoring one model over another, it is clear from these statements that the new Einstein–de Sitter (EdS) model was expected to play an important phenomenological role in cosmology, namely that of providing a zeroth-order description of the universe without incorporating potential, but difficult-to-determine, curvature effects.

It was also clear from the outset that the EdS model enjoyed a privileged physical status, at least if in the absence of empirical evidence model selection is to be guided by Ockham's razor. For instance, in a 1933 influential review, Robertson described the model as "*extremely simple*", emphasizing its minimal assumptions and the avoidance of unnecessary complications (Robertson 1933, p. 78). The differential equations are exactly solvable in closed form, just as analytical are most of the characteristic formulae of the model such as those that make it possible to calculate distances, photon flight times, etc. In particular, by suppressing two major undetermined parameters—the cosmological constant and spatial curvature—the EdS model predicts a simple analytical relationship between cosmic expansion (more rigorously the scale factor of the cosmic metric) and the mean matter density, thereby allowing for a transparent theoretical interpretation and straightforward confrontation with observational data. Interestingly for our discussion, Zwicky himself—independently of Robertson's judgment—recognized the appeal of the recently published EdS model, referring to it as "*the simple Einstein-de Sitter theory*" as early as January 1933 in his cosmic rays paper.

Undoubtedly, the EdS model had more physical than mathematical appeal, as, unlike models of closed curvature, it did not shy away from the notion of an `infinite in act', being unbounded in both space and time. And indeed, from a mathematical standpoint "*compared with the closed universe the open one of Einstein-de Sitter appears to be dulled and uninspired*'' (Infeld, 1949, p.495-496). In any case few theorists and astronomers seemed perturbed by the proposal of an infinite, euclidean geometry for the cosmos, perhaps an indication that they immediately recognized the EdS model as a useful interpretative framework rather than a realistic description of the universe. And indeed, the EdS solution soon became the prototype 'big bang' model for much of the 20th century (Nussbaumer & Bieri, 2009, p.152 ; O'Raifeartaigh, O' Keeffe & Mitton, 2021).

It is in this context of emerging theoretical and observational interests—aimed at investigating the universe's matter content and connecting it to the observed cosmic expansion rate—that Zwicky steps in. One can speculate that for an astronomer so attached to an explanation in terms of simple and known physics of cosmological observations, the EdS models marks a turning point in his appreciation of the general relativistic theories of the universe. He may have been surprised that a theory so far from being intuitive, which makes unconventional predictions, could indeed result in such a simple and compelling cosmological model. After all, a flat, Euclidean model for the geometry of space is the choice we are most familiar with and to which our intuition is most accustomed because it is 'visualisable and depictable'. Indeed, in a book of 1957, where he considers and gauges the hypotheses of an expanding and non-expanding flat universe (Zwicky, 1957, p.171), Zwicky presents the EdS proposal as a "*discussion of what can be considered the simplest possible model of an expanding universe*" (ibid., p.181).

The constant matter density the universe must display in order to have Euclidean spatial sections (about $10^{-28}$ $g/cm^3$) is however much higher than the value it was observationally inferred some years before by Hubble ($\rho=1.5*10^{-31}$ $g/cm^3$) by studying the spatial distribution of visible galaxies. However Hubble commented on his results by stating : "*This must be considered as a lower limit, for loose material scattered between the systems is entirely ignored. There is no means of estimating the order of the necessary correction. No positive evidence of absorption by inter-nebular material, either selective or general, has been found, nor should we expect to find it unless the amount of this material is many times that which is concentrated in the systems*" (1926, p48). As a matter of fact, as previously noted, Zwicky, in reference to Hubble's work, reported the probable interval for the cosmic matter density as $10^{-26}$ $g/cm^3 > \rho > 10^{-31}$ $g/cm^3$.

Neither de Sitter nor Einstein thus considered such a discrepancy as evidence against their expanding but spatially flat model. On the contrary, they claimed that the density needed to flatten space was still within the uncertainties associated with observations, since it "*coincides exactly with the upper limit of the density adopted by one of us*" (p.214). Indeed, in 1930, well before the elaboration of the EdS model, de Sitter had derived from data a matter density value much higher than Hubble's one (de Sitter, 1930). De Sitter obtained his estimate ($\rho=2\cdot10^{-28}$ g/cm$^3$) by multiplying the average number density of galaxies in the universe (about a galaxy in a volume of (0.324 Mpc)$^3$ times the estimated mass of a galaxy. The decisive difference in the Hubble and de Sitter calculations lies in the value adopted for the typical mass of a galaxy. While Hubble assumes m$_{galaxy}$≈2.6·10$^8$ solar mass in his analysis, de Sitter adopts m$_{galaxy}$≈10$^{11}$ solar mass[8]. De Sitter borrows this estimate from the works of Oort (1927$_a$, 1927$_b$) which, unlike Hubble's, "*includes, of course, all gravitating masses: faint and dark stars, interstellar matter, dark nebulae, globular clusters, etc.*" (de Sitter, 1930, p.171). It is clear to de Sitter that the difference between Hubble's calculated luminous matter density and his own could be due entirely to the presence of this dark matter. And the abundance of this component is astonishing: the resulting ratio $\frac{dynamic\ mass}{luminous\ mass}$ is about 385.

Interestingly, in a footnote to a 1931 article, de Sitter remarked (p. 142): "*Hubble derived in Mt. Wilson Contributions No. 324 (1926) a value of $1.5 \times 10^{-31}$ as a lower limit. The value $2 \times 10^{-28}$ adopted by me for convenience in [de Sitter, 1930] was considered as too large by many astronomers.*" Even more notably, in their 1932 paper, Einstein and de Sitter commented on this comparatively large estimate of the matter density parameter, observing that it "*depends on the assumed masses of these nebulae and on the scale of distance, and involves, moreover, the assumption that all the material mass in the universe is concentrated in the nebulae. It does not seem probable that this latter assumption will introduce any appreciable factor of uncertainty.*" This comment makes explicit that the method employed by Hubble and de Sitter to estimate the cosmic density accounted only for the mass contained within galaxies. One may therefore ask whether it is reasonable to assume the existence of a substantial intergalactic abundance of unseen matter. Einstein and de Sitter, however, inclined toward the opposite view.

We may finally note that Einstein himself discussed the physical viability of the EdS cosmological model in a 1932 paper, where he remarked that, for this model to be substantiated, the mean density of the universe should be of "the order of magnitude of *$10^{-28}$g/cm$^3$* (…) which is not inconsistent with astronomers' estimates" (Einstein, cited in O'Raifeartaigh *et al.*, 2015, p. 16). However, Einstein did not reference any specific observational work or result to support this numerical estimate.

Attractive and simple as it is, in light of the discrepancy between the Hubble and de Sitter estimates, the EdS cosmological model lacks the support of robust and decisive observational evidence. This presents itself as a potential conflict and thus interesting opportunity for an observer like Zwicky : should one adopt Hubble's lower estimate or de Sitter's higher one – the value that would provide empirical support for a Euclidean, expanding cosmological model? Furthermore, one may question whether certain forms of matter might remain unaccounted for within the classical methodological approaches used to determine the density parameter. Zwicky's key insight lay in recognizing the possibility of addressing this challenge by analyzing the dynamics of the Coma cluster, thereby providing a new estimate of the cosmic abundance of matter through a method entirely independent of that employed by Hubble and de Sitter –one that allowed for the detection of mass potentially residing outside of galaxies.

---

[8] The de Sitter value is about the value accepted today for the total mass of a typical galaxy.

## 4 –    Zwicky's 1933 paper

The preceding discussion illustrates the context in which Zwicky's research developed and matured as well as his scientific interests, opinions and motivations in 1933. We now focus on the structure and content of his seminal article *The redshift of extragalactic nebulae (1933a)*.

In its abstract, the article presents itself as a generic review of the various techniques used to study extragalactic nebulae and the different theories proposed to explain the intrinsic nature of galaxy redshift. Curiously, neither the method, the innovative application of the virial theorem to a cluster of galaxies, nor the cosmic phenomenon analyzed, the dynamics of the Coma cluster, nor the conclusion, the inferred existence of large amounts of dark matter are mentioned here. On the contrary, the fact that the redshift is a powerful new cosmological observable that will be instrumental in studying cosmic rays is emphasized as the most innovative result of the study "*it will be indicated to what extent the redshift promises to become of importance for the study of cosmic rays.*"

The absence of any reference to the dynamical analysis of Coma or to dark matter in both the title and abstract of Zwicky's paper – sections typically reserved for highlighting the most original and innovative aspects of a work – clearly indicates that the interest of the paper extended well beyond the specific astronomical analysis at hand. That case, of course, was itself highly significant: the first known attempt to apply the virial theorem to a cluster of galaxies. Yet Zwicky's primary concerns were of a broader, explicitly cosmological nature, situated at the intersection of several emerging domains: a newly accessible observable (redshift), the new theoretical framework capable of explaining it (general relativity), and recently discovered astrophysical phenomena (cosmic rays). Furthermore, the absence of the term "dark matter" underscores, as argued in Section 3, the relatively low priority accorded to this concept within the astronomical discourse of the period, despite its latent significance.

The first part of the 1933 article, specifically its Section 2, focuses on analyzing the robustness of the techniques used to determine distances between galaxies. In this regard, it should be noted that Zwicky is particularly interested in the problem of the absorption and scattering of light signals by an invisible intergalactic medium. This problem, in Zwicky's opinion, could hinder the determination of the structure of the universe and in particular may prevent answering "*the first and foremost [….] question whether the distribution of nebulae over space is uniform or not*". Einstein's hypothesis about the uniform spatial distribution of matter (later called by Milne the Cosmological Principle) is the cardinal assumption on which the relativistic expanding models of the universe rely. Interestingly in the introduction Zwicky states "*By and large, the extragalactic nebulae are distributed uniformly over the sky and, as has been demonstrated, are also distributed uniformly in space (…). We must not fail to mention that the above conclusions are only valid in case that absorption and scattering of light in space may be ignored……. In view of the fact that gases and clumps of dust can be proven to exist in interstellar space of our system [Milky Way], it would nevertheless be of great importance to have an independent proof of the transparency of intergalactic space, and to show that it is not the curvature of space, combined with absorption and scattering, that would feign a uniform distribution of the nebulae*".

We note two elements of great interest. Observations, according to Zwicky land support to one of the funding pillars of relativistic cosmological models, the hypothesis that the spatial sections of the universe are maximally symmetric. Second, the issue of dark matter is indirectly introduced through evocation of the potential problems associated with the existence of dust or gas clouds that absorb or scatter luminous signals passing through them. Such a form of nonluminous matter is an obstacle to a proper assessment of the curvature of the spatial surfaces of the universe, which are immediately considered as a real geometrical possibility by Zwicky and not as an abstruse mathematical consequence of the theory of general relativity.

In Section 3, Zwicky goes on to discuss the properties of the observed redshift of galaxies and set up the goal of discussing its nature i.e. what is the physical agent that causes the shift in the spectra of galaxies. He states "*there are presently two general suggestions. The first includes all theories of cosmological character, which are based on the theory of relativity. The second one assumes an interaction of light with matter in the Universe* ". The latter is a hypothesis (later called the tired light hypothesis) put forward by Zwicky in 1929. However, the high transparency of the intergalactic medium, suggests that the loss of energy of the photon cannot be justified in terms of known physical mechanisms ruling the interaction of matter and light (such as the Compton or Raman effects[9]). As a consequence, Zwick was led to speculate about the existence of an ad hoc mechanism, a gravitational friction phenomenon experienced by photons when propagating in a material substratum, as a possible explanation for the redshift, admitting that studies to prove this conjecture were still ongoing.

Both the cosmological interpretation of redshift, as due to the expansion of the universe, and the tired-light hypothesis are not deemed satisfactory by Zwicky because *"all have been developed on extremely hypothetical foundations, and none of these has allowed to uncover any new physical relationship"*.

Interestingly, however, in commenting on the cosmological interpretation of redshift, Zwicky acknowledges that the EdS model, with which *"these two researchers [Einstein and de Sitter] have temporarily given up the existence of an overall curvature of space"* produces many interesting predictions although none of them have yet been confirmed or refuted by observation. The most surprising and decisive prediction concerns the amount of cosmic matter that is distributed in the expanding EdS universe

> *"An expansion of 500 km/s per million parsecs, according to Einstein and de Sitter, corresponds to an average density $\rho \simeq 10^{-28}$ g/cm$^3$. On the basis of observations of self-luminous matter, Hubble estimates $\rho \simeq 10^{-31}$ g/cm$^3$. It is naturally possible that luminous matter and dark (cold) matter together lead to a significantly higher density, and the value of $\sim 10^{-28}$ g/cm$^3$ therefore does not appear unreasonable."* (1933, p.122)

Although Zwicky is often very critical about relativistic models of the universe, he does not use this apparently large discrepancy of about three order of magnitude between the density of luminous matter ($\rho \simeq 10^{-31}$ g/cm$^3$ ) and the density of matter as an argument against the EdS model. The reason is that it is well understood at the time that the luminous matter density is only a lower limit to the density of matter in the universe. For Zwicky, it therefore seems "naturally possible" ("*natürlich möglich*") that by adding the contribution of cold dark matter[10] to Hubble's estimate, a density value close to that predicted by the EdS model could be obtained. As we saw above, this belief is also supported by the fact that both the tired light hypothesis and the model needed to explain the origin of cosmic rays require the presence of large quantities of non-luminous matter.

---

[9] The Raman effect is a phenomenon in which an incident photon interacts with a molecule, exchanging energy with its vibrational or rotational modes, and is scattered with a frequency different from that of the incident photon.

The exact amount of matter distributed in the universe (which , incidentally, no curved model is able to predict) is not the only prediction of the EdS cosmological model. The model also makes a number of other predictions about potentially relevant new physical phenomena, a fact recognised by Zwicky as one of the model's undeniable strengths. For example, the EdS theory predicts that *"for large distances the redshift should increase stronger than linearly with the distance. On the basis of the previous observational material, it is unfortunately not possible as yet to prove this important conclusion."* The EdS model *"also leads to certain conclusions regarding the distribution of brightness levels, number of nebulae, diameter, etc., as function of distance, which however have not yet been proven."*

*Scheinbare Geschwindigkeiten im Comahaufen.*

$$v = 8500 \text{ km/sek} \qquad 6900 \text{ km/sek}$$
$$7900 \qquad\qquad 6700$$
$$7600 \qquad\qquad 6600$$
$$7000 \qquad\qquad 5100 \, (?)$$

*Figure 1 : Apparent velocities of galaxies in the Coma cluster. From (Zwicky, 1933).*

These cosmological predictions await empirical verification, and it is this exciting prospect that motivates Zwicky. He recognised that if redshift – as suggested by relativistic cosmological theories – serves as an indicator of recession velocity, and thus carries dynamical information, it could be used to estimate the abundance of matter in the universe by a method entirely independent of those employed by Hubble and de Sitter. Zwicky's key insight was to extend the application of the virial theorem to clusters of galaxies – that is, to gravitationally bound systems far larger than the Milky Way, beyond the range to which the method had previously been applied. The advantage of applying the theorem on such cosmological scales was evident: rather than merely detecting the gravitational contribution of dark matter within galaxies, this approach could uncover the possible influence of non-luminous matter dispersed throughout intergalactic space.

---

[10] Although Zwicky claims that dark matter is cold, as it is even today largely assumed, the adjective is not used to indicate that dark matter decoupled from the primordial thermal bath after becoming non relativistic (the modern understanding), but simply to mean that dark matter is not emitting in the visible portion of the electromagnetic spectrum.

In presenting his approach, Zwicky explicitly states the two hypotheses on which the virial analysis rests: first, that the cluster is in stationary equilibrium (i.e. neither globally collapsing nor expanding); and second, more crucially for our discussion, that matter is uniformly distributed within the cluster. (A perfectly uniform mass distribution is not strictly required for the virial theorem to hold. However, the weaker condition of spherical symmetry is usually assumed to make the analysis tractable and is sufficient for applying the theorem in the form used by Zwicky)**.** By emphasizing this point, Zwicky highlighted that the reconstruction via the virial theorem makes it possible to infer the total binding mass of the cluster—not merely that confined within its luminous components, i.e., the galaxies themselves. This constituted a major step beyond the "mass counting" approach of Hubble and de Sitter, allowing for an assessment of whether intergalactic matter contributes to the overall mass budget.

The central part of the paper thus focuses on Zwicky's calculation of the velocity dispersion of Coma cluster galaxies, based on their redshifts, and its interpretation within the virial theorem framework. The inferred dispersion of these velocities, called by Zwicky the "medium-sized Doppler effect", turned out to be quite significant, about 1 000 km/s. This is in striking contrast with the theoretically predicted velocity dispersion this structure should display. Indeed, by making reasonable assumptions based on available observations, i.e. that *"the cluster has a radius R of approximately one million light years (equal to 10 24 cm) and contains 800 individual nebulae each of a mass of $10^9$ solar masses"* Zwicky concludes that the velocity dispersion should be as low as 80 km/s.

Zwicky comments his finding as follows:

> *"In order to obtain, as observed, a medium-sized Doppler effect of 1000 km/s or more, the average density in the Coma system would have to be at least 400 times greater than that derived on the basis of observations of luminous matter [This would be in approximate accordance with the opinion of Einstein and de Sitter...]. If this were to be verified, it would lead to the surprising result that dark matter exists at a much greater density than luminous matter" (1933a, p.123)*

So the surprising fact is not the existence of the dark matter, but rather its abundance (the ratio $\frac{dynamic\ mass}{luminous\ mass}$ is about 400). Even more surprising is that the inferred abundance of the dark matter is exactly what is needed to flatten out the spatial hypersurfaces of the cosmological spacetime, at least according to the relativistic EdS model. Indeed as Zwicky points out the value for the ratio between total and luminous mass is very close to that inferred by de Sitter (385).

Incidentally, we note that later in his 1957 book, Zwicky re-determined the density of the Coma cluster based on thermodynamical considerations (the Lane-Emden equation), obtaining values much higher than the EdS expectation. "*Our resulting estimates for the average densities in the cluster are thus $\rho_{(cluster)} > 1.4 \cdot 10^{-26}$ g/cm$^3$ and for the outskirts and the regions between clusters $\rho(intercluster) > 5 \cdot 10^{-27}$ g/cm$^3$. We conclude that the data available so far do not check the Einstein - de Sitter relation*" (Zwicky 1957, p.178).

If redshift is a measure of the recession velocity of galaxies, as advocated by the relativistic models of cosmology, then the large velocity dispersion observed in Coma is not at all an anomaly. A theory is already out there which predicts such a surprising value. Zwicky inferred amount of dark matter is indeed what is demanded if the universe is correctly described by the EdS cosmological model.

It is Zwicky himself to remind us that however appealing this fact might seem, the cosmological interpretation of the redshift effect may not be the last word on the subject. Indeed, in the final section of the paper Zwicky focuses on the issues posed by modelling cosmic rays as being of extragalactic nature, i.e. as originating at great distances from the Earth. The Zwicky 'discover' of a dark matter density which is two order of magnitude larger than the luminous density, might help in justifying the observed cosmic abundance of cosmic rays, as Zwicky himself predicted a few months before. However, Zwicky seems now puzzled by a new issue. Because of the large distances they need to travel, cosmic rays "*would arrive at Earth with very reduced energy as a result of the redshift. If for example the redshift were consistently proportional to distance, then light quanta from a distance greater than 2000 million light years would reach us with zero energy.*"

Interestingly the fact that the EdS model predicts a strong deviation from a linear redshift –

distance relations for large redshifts cannot help in solving the riddle. Indeed the EdS model requires the redshift to increase faster than the distances, contrary to what, according to Zwicky, is needed to support the hypothesis of an extragalactic origin of cosmic rays. Thus, as anticipated in the abstract of the paper, there is still room for investigating possible alternative explanations of the redshift of galaxies.

## 5 – Is Zwicky's dark matter an *ad hoc* hypothesis ?

The traditional interpretation of Zwicky's 1933 findings frames the postulation of dark matter as a hypothesis introduced to account for an unforeseen and puzzling anomaly. As Sanders (2010, p. 13) recounts, "*Zwicky looked at radial velocity measurements (…) and he noticed something quite striking: the galaxies seemed to be moving too fast for the amount of visible matter in the cluster.*" Similarly Bonnet-Bidaud and Lepeltier (2021, p. 221) note: "*To eliminate the anomaly, it was indeed sufficient to postulate the existence of a large quantity of non-visible matter.*" Dark matter hypothesis was initially relegated by influential astronomers largely to the status of "*heresy*" at worst (Biviano, 1997) and of a far-fetched explanation at best. For example, Holmberg characterized this hypothesis as anomalous within the then-accepted paradigm, a view he expressed by stating: "*It does not seem possible to accept the high velocities as belonging to permanent cluster members, unless we suppose that a great amount of mass—the greater part of the total mass of the cluster—is contributed by dark material distributed among the cluster members—an unlikely assumption*" (1940, p. 220). Later critiques echoed similar reservations: de Vaucouleurs argued that "*unless one is prepared to make wild hypotheses outside the realm of verification by direct observation [...] the 'hidden-mass' hypothesis must be ruled out*" (1960, p. 593). Also in more recent epistemological analyses, Zwicky's 'hidden mass' is presented as a sudden ad hoc hypothesis without being accorded even the minimal ontological status of 'hidden luminous matter.' For example Lazutkina (2017, p. 7) observes that "*Fritz Zwicky studied the galaxy cluster Coma and found that the visible mass is insufficient to account for the observed gravitational effects. (…) The observations did not fit the theory, so a hypothetical object was introduced.*"

On this reading, the dark matter hypothesis functions as an auxiliary construct, formulated specifically to account for an unexpected anomaly within the prevailing paradigm of the time. It is therefore crucial to assess whether Zwicky's proposal genuinely meets the criteria for being qualified as *ad hoc*. The term *ad hoc* is also largely adopted to qualify the introduction in cosmology of *the modern dark matter hypothesis* — that is, matter not accounted for within the Standard Model of particle physics. As Merritt succinctly states, *"the dark matter hypothesis is ad hoc in the sense of not satisfying the heuristic of the program: the hypothesis was added purely in response to unanticipated observational facts"* (2020, p. 34). Assessing the appropriateness of such a qualification lies beyond the scope of the present analysis, which is concerned exclusively with Zwicky's dark matter hypothesis.

The notion of the ad hoc hypothesis occupies a pivotal position within the philosophy of science. For certain thinkers—most notably Kuhn and Feyerabend—it constitutes an indispensable element of scientific practice and theory adjustment, whereas for others, such as Karl Popper[11], it represents a 'conventionalist stratagem' because it is introduced to 'save' theories, and if a accepted theory is protected by the introduction of ad hoc hypotheses when it is challenged, it becomes irrefutable.

Although the role of the status of the ad hoc hypothesis is controversial, it is nevertheless sufficient to share a few elements of definition in order to examine whether Zwicky's postulation of a substantial quantity of dark matter satisfies shared criteria that allow it to be classified as an ad hoc hypothesis. Between the positions of Kuhn and Popper, Jarrett Leplin combines some of these elements to offer us a broad and nuanced view of what an ad hoc hypothesis is. He thus identifies five analytic criteria :

*"An hypothesis H introduced into a theory T in response to an experimental result E [in our case it is an observation] is ad hoc if and only if:*

>   *(1) E is anomalous for T, but not for T as supplemented by H.*
>   *(2) E is evidence for H, but*
>       *(a) no available experimental results other than E support H,*
>       *(b) H has no application to the domain of T apart from E.*
>       *(c) H has no independent theoretical support.*
>   *(3) There are sufficient grounds neither for holding that H is true nor for holding that H is false.*
>   *(4) H is consistent with accepted theory and with the essential propositions of T.*

---

[11] We could also mention Carl Hempel *Some genuinely testable theories, when found to be false, are still upheld by their admirers—for example by introducing ad hoc some auxiliary assumption, or by reinterpreting the theory ad hoc in such a way that it escapes refutation.*' (1966, p.37)

> *(5) There are problems other than E confronting T which there is good reason to hold are connected with E in the following respects :*
>> *(a) these problems together with E indicate that T is non-fundamental,*
>> *(b) none of these problems including E can be satisfactorily solved unless this non-fundamentality is removed,*
>> *(c) a satisfactory solution to any of these problems including E must contribute to the solution of the others." (1977, p.336-337).*

Failure to satisfy any one of these criteria disqualifies the hypothesis from bearing the ad hoc designation. We argue that the first two are sufficient for us to dispute this status.

Criterion (1) is not satisfied. As we have argued, the gravitational effects of non-luminous matter were neither unforeseen nor unprecedented; they were anticipated and, in fact, observed by several of Zwicky's contemporaries and predecessors. Within both classical and relativistic frameworks, there is nothing anomalous in positing that the dynamics of massive bodies can reveal the underlying mass distribution more fully than observations limited to luminous matter. Indeed, it is theoretically natural—within any gravitational theory (T)—to expect that gravitational interactions reflect the total mass, whether luminous or not. In this light, the true anomaly would have been the reverse: if the amount of visible matter had exceeded the mass inferred from dynamical considerations. Thus, Zwicky's dark matter hypothesis does not meet the criterion of responding to an anomaly in the relevant theoretical context.

Nor are criteria (2.a) and (2.b) satisfied, given the extensive and substantive epistemic integration of Zwicky's dark matter hypothesis with other theoretical and observational elements, as our analysis has demonstrated. Far from being an isolated or ad hoc addition, the hypothesis served as a unifying thread that wove together multiple, seemingly disparate astronomical phenomena of the time. Zwicky envisaged dark matter as a conceptual bridge linking Oort's findings on the unexpectedly high velocity dispersion of stars perpendicular to the galactic plane—dynamics not attributable solely to visible matter—with the Euclidean spatial geometry implied by de Sitter's cosmological model, as well as his own conjectures regarding the origins of cosmic rays. Under the assumption of a substantial quantity of unseen mass, these phenomena form a coherent explanatory network, thereby undermining any claim that the hypothesis lacks independent support or systematic connection to broader theoretical commitments.

So, we are led to reject the view that Zwicky's hypothesis was merely a convenient fix formulated to explain just the one phenomenon (the large velocity dispersion of the galaxies in the Coma cluster) under investigation. We instead portray it as a legitimate and potentially progressive scientific proposal. This hypothesis forged in a network of *positive connections*[12] (Israel-Jost, 2015) between existing theoretical proposals and empirical observations. In this way, it contributes to the coherence of the epistemic context. It is in this latter, integrative sense that the introduction of the dark matter hypothesis should be understood.

Likewise, the remaining propositions – particularly (3) and (5) – fail to capture the genuine epistemological content of Zwicky's hypothesis. As we demonstrate, the dark matter hypothesis cannot be reduced to a mere stratagem devised to preserve a conventional theory or to render empirical evidence compatible with it.

According to Sanders, "*the reconciliation of astronomical observations with Newtonian dynamics is (...) the original motivation" for the Zwick's dark matter hypothesis*[13]" (2010, p.12). This narrative presents Zwicky as being fatally locked into a Newtonian paradigm. We show, on the contrary, that the fact that Zwicky employed Newtonian dynamics to study the motion of galaxies in the Coma cluster should not be interpreted as such, nor should the introduction of the dark matter hypothesis be viewed as an attempt to preserve Newtonian gravity. Rather, his choice is best understood as a pragmatic one: in the low-velocity regime characteristic of the Coma galaxies, the predictions of Newtonian and relativistic gravitational theories are effectively indistinguishable. Thus, Newtonian dynamics served as an adequate and operationally convenient framework for analyzing the data, without implying allegiance to its fundamental correctness. Indeed, as we have argued, the dark matter hypothesis was not a conservative move to preserve the classical framework, but rather an innovative way to transit from the standard Newtonian theory of gravitation to Einstein's relativistic theory.

That the research program of Zwicky was centrally concerned with challenging the established Newtonian paradigm is evident in his conception of the large-scale distribution of galaxies as an ideal laboratory in which to "*test the validity of the inverse square law of gravitational forces*" (Zwicky 1937).

---

[12] Vincent Israël-Jost writes: "Coherence must imply a positive connection between the beliefs of the system and cannot be based on the mere absence of conflict" (2015, p.100).
[13] And also for the modern hypothesis.

In this light, we reject the common analogy that likens the dark matter episode in Zwicky's work to Le Verrier's prediction of Neptune. Le Verrier, in his time, invoked the ad hoc existence of an as-yet-unseen planet to account for irregularities in the orbits of known celestial bodies—thereby preserving the integrity of classical Newtonian gravity (Worrall, 2005; Schindler, 2018). Zwicky, by contrast, did not seek to shield an established theory from anomalous data; rather, he sought to provide empirical support for newly emerging interpretive frameworks—not only his own tired-light redshift hypothesis and model for the intergalactic origin of cosmic rays, but also Einstein's general relativity—through the postulation of an invisible mass. Where Le Verrier's move was conservative, seeking to restore coherence within a trusted paradigm, Zwicky's was exploratory, extending trust to a novel theoretical vision in search of confirmation.

## 6 On the durability of the *ad hoc* narrative

Having deconstructed the conventional narrative that portrays dark matter as an ad hoc hypothesis devised to preserve an established paradigm, we now turn to the question of its enduring popularity. One significant factor lies in the growing disconnect between the communities of theorists and observers—a rift that would only deepen in the years following Zwicky's pioneering work. In contrast to Zwicky's approach—rooted in his rare command of both physicists' and astronomers' analytical tools, and guided by a commitment to testing, challenging, or even advancing cosmological theory through observation—the methodology shifts markedly in the work that follows. In the work of, for example, Sinclair Smith, in which Zwicky's results are confirmed through a similar virial analysis of the Virgo cluster, no more connection is made between the prediction of the theory (the EdS model) and the amount of matter observed in the universe. The umbilical cord that tied cosmological theory predictions to its empirical confirmations had been prematurely cut. The problem of dark matter would soon be relegated to becoming only a problem of celestial mechanics applied to galaxy clusters, the test-bed for Newtonian laws of motion and not for the predictions of the theory of general relativity, a problem of astrophysics and not of cosmology. That Zwicky failed to convince the community that dynamical studies of clusters could provide observational evidence for a cosmological model is corroborated by the fact that his article—published in German in a little-known Swiss journal—was not cited for the following 25 years[14].

We raise the question of whether the widespread perception of Zwicky's hypothesis as *ad hoc* has, in turn, shaped alternative characterizations or interpretations of what constitutes an ad hoc hypothesis. For instance, Mary Hesse observed (Hesse, 1961, p.228) that such hypotheses almost invariably invoke unobservable entities or mechanisms. This perspective may partly explain why the dark matter hypothesis has so often been considered ad hoc, given the presumed unobservability of dark matter. Although this view was later superseded by Jarrett Leplin's more formal and influential definition, Hesse's interpretation still surfaces in many naïve or informal understandings of the term. Conversely Hunt (2012) has challenged the idea that unobservability alone suffices to render a hypothesis ad hoc. After all, what is unobservable today may become observable tomorrow, through theoretical refinement or technological advancement—as might well be the case with dark matter. We would add that the notion of observability in physics is far from straightforward. It cannot be reduced to the mere activation of a detector or the stimulation of the observer's eye. Rather, what counts as observable object is deeply shaped by the theoretical framework through which phenomena are interpreted. Whether the curvature of space or electromagnetic waves are deemed observable depends on our willingness to recognize their effects—their capacity to leave measurable traces or induce changes in other systems—as legitimate forms of observation. In this light, dark matter, though not directly detected, is already "indirectly observed" through its gravitational influence on the motion of both massive and massless particles.

Additionally it seems to us that the narrative of the ad-hoc hypothesis is also entertained by proponents of modified gravity theories as an alternative explanation for dark matter. For example Milgrom (2020), presents dark matter as a new form of 'ether'[15] the prototype of ad hoc and poorly motivated hypothesis that recurrently punctuated the history of physics and astronomy. Presenting dark matter as an a posteriori hypothesis, introduced ad hoc to preserve the accepted paradigm, is functional to flog conventionalism, the attitude with which mainstream scientists would hinder the emergence of alternative theories.

---

[14] Today, it has over 1500 citations according to NASA/ADS and over 4000 according to Google Scholar
[15] Which Sander defines as 'an environment that is necessary to make observations consistent with the expectations of existing theory' (2010, p.6)

A caveat is in order. The arguments and conclusions advanced here regarding the epistemological status of Zwicky's formulation of dark matter cannot be straightforwardly extended to the modern dark matter hypothesis—conceived not merely as non-luminous but also as non-baryonic, that is, as composed of entities lying beyond the elementary constituents of the Standard Model of particle physics. It is worth noting, however, that Man Ho Chan (2019) has recently contended that even in its contemporary formulation, the dark matter hypothesis should not be regarded as having been introduced in an *ad hoc* manner.

## 7 The epistemic standing of the *ad hoc* qualification

Some authors have questioned the relevance of distinguishing a special status for hypotheses "expressly intended for a purpose". Hunt notes that "*surely all scientific hypotheses are generated for some purpose, to explain something or another that has been observed about nature*" (2012, p.2). This might lead one to think that judging the *ad hoc* status or not of a hypothesis seems only possible *a posteriori*. Hempel supports this idea: "*with the benefit of hindsight, it seems easy to dismiss certain scientific suggestions of the past as ad hoc hypotheses, whereas it might be quite difficult to pass judgement on a hypothesis proposed in a contemporary context*" (1966, p.30). This could be illustrated by the fact that the discovery of Neptune is nowadays presented as a great prediction and the planet Vulcan as an *ad hoc* hypothesis. However, such a distinction was impossible in Le Verrier's epistemic context. When they were still hypotheses, they had the same epistemological status.

According to these authors, to say that a model is or is not based on ad hoc assumptions is just an aesthetic choice. Hunt even holds an extreme conclusion: "*Considering all this, it seems to me that hand-wringing by historians, philosophers, and scientists over whether or not a hypothesis is or was ad hoc (and therefore "illegitimately proposed") is energy wasted. If what is ad hoc depends on an aesthetic sense (and is thus largely in the eye of the beholder) and, furthermore, can change after the fact, I suggest that the usefulness of the term "ad hoc hypothesis," if it ever had any, has come to an end. At the end of the day there seem to be no ad hoc hypotheses and no non–ad hoc hypotheses, only hypotheses—full stop.*" (ibid., p.13)

Holton agrees, arguing that the assignment of *ad hoc* status is based on the "feeling of the scientist" (1988, p.326)[16]. This would lend itself quite well to the controversy between dark matter and modified gravity theory. Proponents of the modified theory of gravity MOND, proposed to banish dark matter from the cosmological scenario, believe that the latter is an ad hoc hypothesis because it was introduced after observing a discrepancy between the dynamically inferred mass of galaxies and that inferred from luminosity. Proponents of dark matter, on the other hand, believe that it is the MOND model that is ad hoc because the model was explicitly designed to explain the observed dynamics of galaxies and does not allow for prediction or explanation of other phenomena than galaxy rotation curves.

---

[16] Although he doesn't see it as a problem.

The fact that two communities accuse each other of employing *ad hoc* explanations, is not sufficient to conclude that the *ad hoc* status is useless. It only emphasizes the need for a clear and rigorous redefinition of what ad hoc means.

In this spirit, one might adopt the least contentious construal of the term *ad hoc*: a retrospective label applied to hypotheses that have failed to secure a durable standing within the scientific enterprise. While not endorsing this view, we nevertheless take inspiration from its minimalism in order to advance a constructive proposal for diagnosing when a hypothesis should properly be regarded as *ad hoc*.

Rather than abandoning the distinction between ad hoc and non–ad hoc hypotheses, as Hunt explicitly recommends, we propose instead to assess the *quality* of hypotheses along a scale that measures, at a given time, their degree of connectivity with the whole network of physical phenomena. At the lowest point of this scale lies the *ad hoc hypothesis*, defined as a statement devised for a single explanatory purpose and no deeper reach. The mid-range of the scale is represented by the *coherent hypothesis*, which demonstrates a capacity to connect with and support a larger set of theoretical or empirical elements. At the upper end stands the *theoretical cornerstone*, a hypothesis whose integration and robustness elevate it to the status of a foundational element within scientific theory. This framework thus traces a continuum of epistemic virtue, ranging from the merely provisional, through the working, to the firmly established hypothesis. On this view, the epistemic merit of a hypothesis is thus a dynamical variable that evolves in time according to the progress of understanding and discoveries. To give concrete examples an ad-hoc hypothesis would be the Einstein's cosmological constant which was formulated *ad hoc*, to enforce the staticity of a spherically symmetric 3D distribution of cosmic matter, to be promptly dismissed as the dynamical nature of spacetime became soon after evident thanks to the observations of Hubble. Curiously, after the observations of the unexpected dimming of light from distant supernovae, it has now re-surfaced on the cosmological stage, to explain the accelerated nature of the expansion of the universe, but having virtually no other explanatory purpose nor effect than on the geometry of spacetime.

Higher up this scale are the hypotheses whose potential for connection we know without it being fully accessible at the moment they are formulated. Pauli 's hypothesis of the existence of the neutrino, postulated in 1930 to explain the β decay without violating the well established conservation laws of energy, momentum and spin, is certainly an *ad hoc* hypothesis. However, since one is potentially able to determine the properties of this particle both theoretically and by experiment, and since this particle intervenes in many other reactions (the transformation of a neutron into a proton[17], the polarisation of the photon emitted following electron capture[18], the oscillation of neutrinos[19], etc.), one cannot consider this hypothesis as intended for a single use.

Zwicky's dark matter provides another illustrative example of this category of hypothesis. As we have emphasized, dark matter possessed multiple explanatory affordances within the epistemic context of the time. Its role as a unifying element across several seemingly unrelated phenomena—the vertical dynamics of stars in the Galactic plane, the matter density anticipated in Einstein's de Sitter model, and the gravitational potential energy of the Coma cluster—suggests that, at least within Zwicky's *Weltanschauung*, it already qualified as a coherent hypothesis.

In this scheme, another hypothesis advanced in cosmology, the so called cosmological principle—namely, the assertion that the three-dimensional spatial sections of the universe are invariant under translations and rotations—may be classified as a cornerstone hypothesis, given its foundational role in the dynamical theory of the universe and its far-reaching implications for a wide range of independent cosmic phenomena (Marinoni, 2025).

## Conclusion

The status of the notion of dark matter prior to 1933—the year conventionally regarded as the concept's birth—is examined in order to contextualize the epistemological landscape in which

---

[17] Discovered experimentally in 1956 by Frederick Reines and Clyde Cowan in a nuclear reactor.
[18] In 1957 and 1958, Maurice Goldhaber, Lee Grodzins and Andrew Sunyar determined experimentally that the spin of the neutrino is antiparallel to its momentum.
[19] Discovered by Masatoshi Koshiba in 1998 using the Super-Kamiokande detector. This discovery earned him a Nobel Prize 4 years later.

Zwicky's article *'On the Redshift of Extragalactic Nebulae'* is situated, and to better frame the ostensibly 'not unexpected' novelty of its conclusion concerning the existence of dark matter.

Furthermore and more interestingly, a critical reading of the article reveals the profound cosmological, and not merely astronomical, motivations that led Zwicky to engage with the problem of dark matter. His results on the abundance of dark matter must be understood in the context of Einstein's emerging relativistic theory of the universe—a framework of which Zwicky was aware and whose technical aspects he had mastered. The large ratio between invisible matter (more precisely, matter not emitting in the optical band) and visible matter that he derived is neither accidental nor surprising. It corresponds exactly to the values required to substantiate and render observationally viable the cosmological model of Einstein and de Sitter.

In this new light Zwicky appears as the quantifier, not the discoverer, of dark matter, the physicist who ingeniously found a way to adapt an existing tool, the viral theorem, to weight the mass of the universe, finding that it matched exactly what was predicted by theories and models on the scientific scene in 1933. Although Zwicky is more legitimately the physicist who solidly and robustly confirmed the existence of dark matter rather than its theorizer, he deserves credit for recognizing that dark matter was a multipurpose concept needed by several disciplines, astronomy, cosmology and high energy physics.

By challenging the characterization of Zwicky's identification of dark matter as ``surprising" or "naïve," and by disputing the claim that it was wholly unanticipated or disconnected from existing scientific frameworks and prior cosmological expectations, we seek to deconstruct the traditional narrative that portrays the dark matter hypothesis as merely ad hoc. The historical reconstruction instead suggests that dark matter should be regarded not as an a posteriori necessity imposed by theory, but rather as an a priori hypothesis—one arguably less audacious than the assumption that all cosmic matter is luminous.

We propose several ways to assess the resilience of the widespread account and to contribute to the debate on whether a hypothesis should be qualified as ad hoc. To that end, we outline a minimalistic interpretation of what *ad hoc* ought to mean in the physical context of hypotheses and introduce a scale for evaluating their quality. From this perspective, dark matter—at least in Zwicky's original formulation—emerges not as an ad hoc postulate, but as a coherent and unifying hypothesis. It was advanced to provide a plausible interpretation of multiple, seemingly unrelated enigmas in the scientific landscape of 1933.

## Acknowledgments

The authors thank Julien Bernard for his comments and discussions.

The authors are grateful to the anonymous reviewers of the *Journal for General Philosophy of Science* for their valuable feedback.

SB completed his PhD thanks to funding from the doctoral contract *Inted-ED* of Aix-Marseille Université.

CM is supported by the French government under the France 2030 investment plan, as part of the Initiative d'Excellence d'Aix-Marseille Université -  A*MIDEX (AMX-19-IET-012).